\documentclass[letterpaper, 10 pt, conference]{ieeeconf}  

\IEEEoverridecommandlockouts                              

\usepackage{graphicx} 
\usepackage{amsmath, dsfont, amssymb, amsthm}
\usepackage{hyperref}
\usepackage{braket}
\usepackage{xcolor}
\usepackage{makecell}
\usepackage{multirow}
\usepackage{cite}

\DeclareMathOperator*{\argmin}{argmin} 

\usepackage{color, colortbl}

\usepackage{titlesec}

\titlespacing\section{0pt}{1pt}{1pt}
\titlespacing\subsection{0pt}{2pt minus 1pt}{2pt minus 1pt}
	
\definecolor{green}{rgb}{0.9,1,0.9}

\newcounter{sideremark}

\title{\LARGE \bf Multiagent Reinforcement Learning for Autonomous Routing and Pickup Problem with Adaptation to Variable Demand
}
\author{Daniel Garces$^{1}$, Sushmita Bhattacharya$^{1}$, Stephanie Gil$^{1}$, Dimitri Bertsekas$^{2}$
\thanks{$^{1}$Daniel Garces, Sushmita Bhattacharya, and Stephanie Gil are with the REACT lab, Harvard University, Boston, MA, USA
(e-mail: {\tt\small [dgarces]@g.harvard.edu}, {\tt\small sushmita\_bhattacharya@g.harvard.edu}, {\tt\small sgil@g.harvard.edu}). }%
\thanks{$^{2}$Dimitri Bertsekas is with the Department of Electrical Engineering and Computer Science
, Arizona State University, AZ, USA (e-mail: {\tt\small dimitrib@mit.edu}).}
\thanks{This work was supported by 
ONR YIP (grant \# N00014-21-1-2714),
and Amazon Research Award.}
}

\begin{document}

\maketitle
\thispagestyle{empty}
\pagestyle{empty}

\begin{abstract}
We derive a learning framework to generate routing/pickup policies for a fleet of autonomous vehicles tasked with servicing stochastically appearing requests on a city map. We focus on policies that 
1) give rise to coordination amongst the vehicles, thereby reducing wait times for servicing requests, 
2) are non-myopic, and consider \emph{a-priori} potential future requests, 
3) can adapt to changes in the underlying demand distribution. 
Specifically, we are interested in policies that are adaptive to fluctuations of actual demand conditions in urban environments, such as on-peak vs. off-peak hours. We achieve this through a combination of (i) an online play algorithm that improves the performance of an offline-trained policy, and (ii) an offline approximation scheme that allows for adapting to changes in the underlying demand model. In particular, we achieve adaptivity of our learned policy to different demand distributions by quantifying a region of validity using the q-valid radius of a Wasserstein Ambiguity Set. We propose a mechanism for switching the originally trained offline approximation when the current demand is outside the original validity region.
In this case, we propose to use an offline architecture, trained on a historical demand model that is closer to the current demand in terms of Wasserstein distance. 
We learn routing and pickup policies over real taxicab requests in San Francisco with high variability between on-peak and off-peak hours, demonstrating the ability of our method to adapt to real fluctuation in demand distributions.
Our numerical results demonstrate that our method outperforms alternative rollout-based reinforcement learning schemes, as well as other classical methods from operations research. 
\end{abstract}

\section{Introduction}

We consider the problem of dynamic multiagent autonomous taxicab routing for rider pickups, a special case of the Dynamic Vehicle Routing (DVR) problem \cite{BERBEGLIA20108}. This problem has relevance to several robotics tasks, including coordinated package delivery ~\cite{choudhury2021efficient} \cite{Arbanas2016} \cite{Zhang2022}, warehouse robot path planning \cite{Emanuelson2022}, autonomous transportation, and on-demand mobility systems ~\cite{Tsao2019} \cite{Guo2022} \cite{Li2022} \cite{Mora2017} \cite{Gueriau2020}. We express the demand as riders waiting to be transported, and assume that taxicabs can transport one rider at a time. We are interested in finding a strategic, cooperative pickup plan for a fleet of autonomous taxicabs to minimize the total wait time of all riders where the number and location of future requests are not known \textit{a-priori}. Obtaining an optimal solution for this problem is intractable since it requires considering multiple scenarios of potential future requests and all relevant taxicab actions at each decision point. This condition results in an extremely large state space and a control space that grows exponentially with the number of agents. 
Hence, finding a competitive sub-optimal solution is crucial.

Several sub-optimal solutions to the taxicab pickup problem explore instantaneous assignment \cite{Bertsekas1979Auction} \cite{Karp1990} \cite{Duan2014}  \cite{Bertsimas2019OnlineVR}, and other routing heuristics, including 2-opt \cite{Croes1958}, local search \cite{LSCP2003}, and genetic algorithms \cite{Mitchell1998}. These methods tend to generate myopic policies due to the lack of consideration for future demand. Sampling-based stochastic optimization methods that consider potential future requests~\cite{LOWALEKAR201871} address this limitation, but at the expense of long computation times due to multistep planning in a large state space. Several learning-based approaches aim to achieve faster computation times. Some authors consider offline trained approximations, such as approximate value iteration\cite{Ulmer2018}, Deep Q-learning \cite{PARVEZFARAZI2021100425}, Deep Q Networks \cite{AHAMED2021227}, and transformer-based architectures \cite{Wu2016Diamond}. Offline trained architectures may fail to generalize to unknown scenarios, not fully represented in the training data. This condition makes them infeasible for deployment in real urban environments with fluctuating demand. Other authors consider online policy evaluation with finite lookahead, including Monte Carlo Tree Search (MCTS) \cite{SiV10}, DESPOT \cite{Despot}, and a Multiple Scenario Approach (MSA) \cite{Bent2004b}. These methods tend to be computationally expensive if no approximations are used.

In this paper we aim to address the lack of generalization of offline trained architectures, and the high policy evaluation time of online methods, by proposing a hybrid planning method with online optimization on top of an offline trained policy approximation. Our method obtains a competitive suboptimal solution to the taxicab routing problem by leveraging (i) \emph{online play}, a lookahead optimization scheme that improves on the results of offline training \cite{bertsekas2020rollout} \cite{Bertsekas2022AlphaZero}, and (ii) an offline approximation switching scheme that endows the system with adaptivity in the face of significant changes in the underlying demand model. We use an offline trained approximation as the base policy for online play for faster computation times compared to simulation-based rollout algorithms. Our offline approximation is implemented using Graph Neural Networks (GNNs)\cite{Scarselli2009} that exploit the topological characteristics of a city environment. We achieve adaptivity for the offline approximations by formulating Wasserstein Ambiguity Sets \cite{Esfahani2015} centered around representative historical demand models. These sets represent regions of validity, or probability spaces in which the offline approximations correctly approximate the rollout based Reinforcement Learning (RL) method~\cite{bertsekas2020rollout} under a particular demand model.

\begin{figure}[ht]
        \vspace{-5pt}
        \centering
        \includegraphics[width=0.30\textwidth]{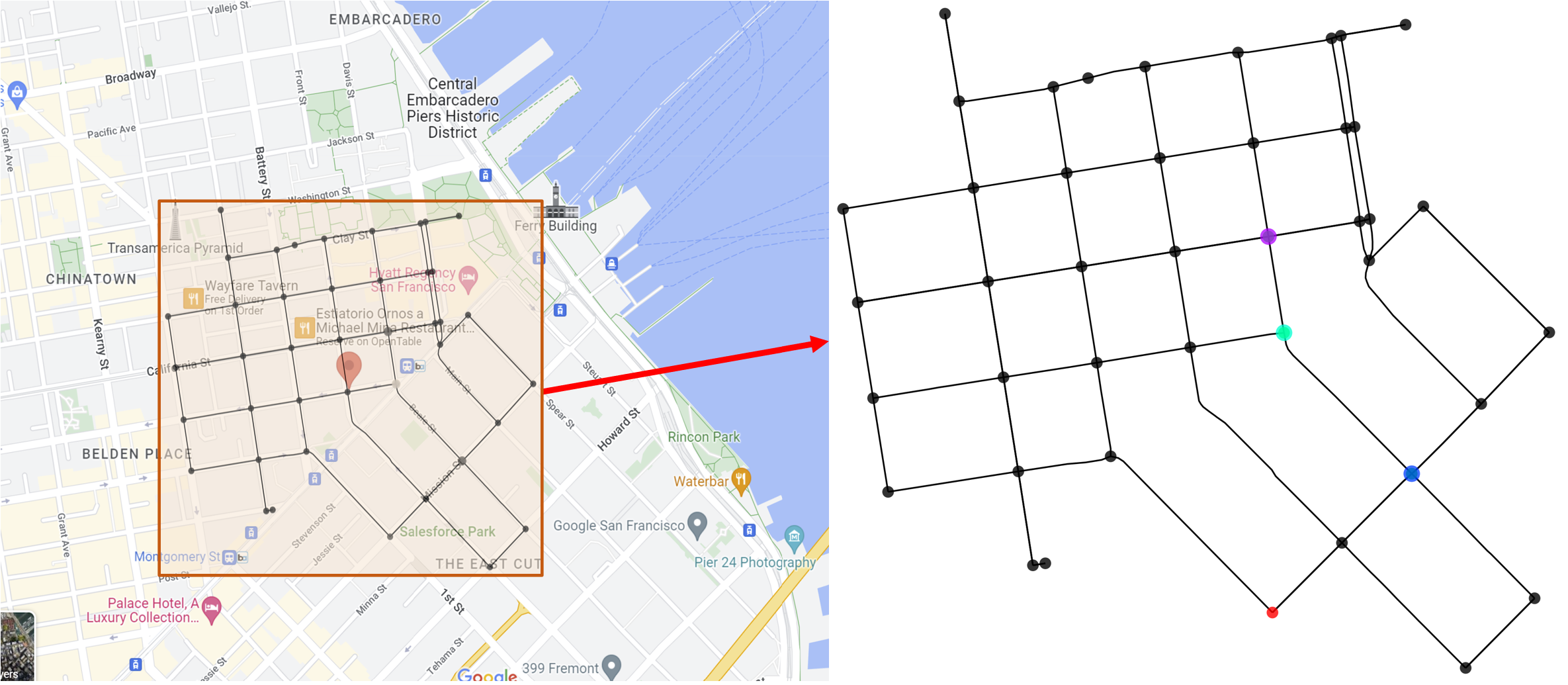}
        \vspace{-8pt}
        \caption{\small{Example of a topological map and corresponding graph preserving one-way streets and traffic direction constraints. 
        }}
        \vspace{-10pt}
        \label{fig:graph/map}
    \end{figure}

The main contributions of this work are as follows: 
1) We formulate the autonomous taxicab routing problem as a stochastic Dynamic Program (DP) 
in such a way that an offline trained GNN successfully approximates a rollout-based RL method. This then allows us to apply online play as an approximate Newton step \cite{Bertsekas2022AlphaZero} to further improve the performance of the learned policy. 
\noindent 2) Our method adapts to fluctuating demand conditions by
switching between offline approximations to maintain the performance improvement of online play. We replace the offline approximation once the current demand is no longer inside the approximation's region of validity.
\noindent 3) We apply our approach on a real taxicab pickup dataset~\cite{epfl-mobility-20090224} (an example map is shown in Fig.~\ref{fig:graph/map}).
We empirically show that our method outperforms one-at-a-time rollout with a simple base policy, and several classical benchmark algorithms operations research (OR).

The rest of the paper is organized as follows: In Sec.~\ref{sec:problem_formulation}, we present our formulation of the multiagent taxicab routing problem. In Sec.~\ref{sec:our_approach}, we present our approach, our application of online play with offline approximation, and our formulation of Wasserstein Ambiguity Sets to quantify regions of validity. Finally, in Sec.~\ref{sec:result} we present numerical results on the San Francisco taxicab pickup dataset~\cite{epfl-mobility-20090224}.

\section{Problem formulation}\label{sec:problem_formulation}
Here, we present the formulation for a multiagent taxicab routing problem as a discrete time, finite horizon, stochastic DP problem. 
In the following subsections, we outline the environment, requests, state and control representations.

\subsection{Environment} \label{subsection_environment}
We assume that taxicabs are deployed in urban environments with a fixed street topology, represented as a directed graph, $\mathds{G} = (\mathds{V}, \mathds{E})$. Here, $\mathds{V} = \{ 1, \dots, n\}$ corresponds to the set of indices for the street intersections in the map of the city with a total of $n$ intersections, and $\mathds{E} \subseteq \{(i, j) | i, j \in \mathds{V} \}$ corresponds to the set of directed streets connecting intersections $i$ to $j$. The set of \emph{neighbors} of intersection $i$ is denoted by $\mathcal{N}(i) = \{j| j \in \mathds{V} , (i, j) \in \mathds{E}\}$.

\subsection{Requests} \label{subsection_requests}
A request $r$ is represented as a tuple $r = (\rho_{r}, \delta_{r}, k_r, \phi_{r})$, where $\rho_{r} \in \mathds{V}$ and $\delta_{r} \in \mathds{V}$ correspond to the closest intersections to the desired pickup and dropoff locations for the request, respectively; $k_r$ corresponds to the time at which the request enters the system; and $\phi_{r} \in \{ 0, 1\}$ is an indicator variable which equals $0$ if the request has not been assigned to any agent, otherwise $\phi_{r} = 1$.
We model the number of requests that enter the system as a random variable $\eta$ with an unknown underlying distribution $p_\eta$, and we denote its realization at time $k$ as $\eta(k)$. $p_\eta$ is fixed for the entire length of the time horizon $N$ and we estimate it using either historical data or data for the last hour of execution of the system, to obtain categorical probability distributions $\Tilde{p}_\eta$ (historical) or $\Tilde{p}_{\eta,c}$ (current), respectively. We denote the set of requests that enter the system at time $k$ as $\mathbf{r_k}$. Here the cardinality of the new request set at time $k$ is $|\mathbf{r_k}|=\eta(k)$. 

We define $\mathbf{\Bar{r}_k}$ as the list of outstanding requests that have not been assigned to any agent till time $k$, such that $\mathbf{\Bar{r}_k} = \{ r | r \in \mathbf{r_t}, \phi_{r} = 0, 1 \leq t \leq k\}$. The pickup $\rho_{r}$ and dropoff $\delta_{r}$ locations of a new request $r$ are determined by underlying probability distributions $p_\rho$ and $p_\delta$, where $\rho$ and $\delta$ correspond to random variables for the pickup and dropoff locations, respectively. These two probability distributions are also unknown a-priori. We estimate the corresponding categorical distributions $\Tilde{p}_\rho$ and $\Tilde{p}_\delta$ using historical data, and 
$\Tilde{p}_{\rho, c}$ and $\Tilde{p}_{\delta, c}$ using the last hour data. The estimated distributions $(\Tilde{p}_\eta, \Tilde{p}_\rho, \Tilde{p}_\delta)$ compose the historical demand model and  $(\Tilde{p}_{\eta,c}, \Tilde{p}_{\rho, c}, \Tilde{p}_{\delta, c})$ compose the current demand model. We discuss the estimation process in Sec.~\ref{sec:demand_model}.

\subsection{State representation and control space} \label{subsection_state}
We assume there are a total of $m$ agents and each agent can perfectly observe all requests, and all agents' locations and occupancy status. The state at time $k$ is $x_k = (\boldsymbol{\nu}_k, \boldsymbol{\tau}_k, \mathbf{\Bar{r}_k})$. Here, $\boldsymbol{\nu}_k = [\nu_k^1, \dots, \nu_k^m]$, is a list of locations for all $m$ agents at time $k$, where $\nu_k^\ell \in \mathds{V}$ is the index of the closest intersection to the geographical coordinates of agent $\ell$. $\boldsymbol{\tau}_k = [\tau_k^1, \dots, \tau_k^m]$, is a list of time remaining in assigned trip for all agents at time $k$, and $\mathbf{\bar{r}}_k$ is the set of outstanding requests at time $k$. If $\tau_k^\ell = 0$, then agent $\ell$ is available and new requests can be assigned to the agent, otherwise $\tau_k^\ell \in \mathds{N}^+$. 

The control space of agent $\ell$ at time $k$ is denoted by $\mathbf{U}_k^\ell(x_k)$. If $\tau_k^\ell = 0$ (agent is available), then $\mathbf{U}_k^\ell(x_k)=\{\mathcal{N}(\nu_k^\ell), \nu_k^\ell , \zeta\}$, where $\zeta$ is a special pickup control that becomes available if there is a request $r \in \mathbf{\bar{r}}_t, 1 \leq t \leq k$, such that its pickup location $\rho_{r}=\nu_k^\ell$. 
If $\tau_k^\ell \neq 0$, then the agent must complete its current trip before picking up a new request. So $\mathbf{U}_k^\ell(x_k)=\{h\}$, where $h$ is the next hop in the shortest path (given by Dijkstra's algorithm) between $\nu_k^\ell$ and the dropoff location $\delta_{r}$ for the agent's assigned request $r$. Since this formulation represents a separable control constraint for each agent, the controls available to all agents at time $k$,  $\mathbf{U}_k(x_k)$, is expressed as the Cartesian product of local control sets $\mathbf{U}_k^1(x_k) \times \dots \times \mathbf{U}_k^m(x_k)$.

\subsection{Stochastic dynamic programming formulation} \label{subsection_cost}
We now present our formulation of the taxicab routing problem as a finite horizon, stochastic DP problem.
The objective is to find a pickup strategy that minimizes the total wait time of requests (in minutes).
The state transition function is denoted by $f_k$, and $x_{k+1} = f_k(x_k, u_k, \eta, \rho, \delta)$, where $x_{k+1}$ is the state at time $k+1$ after application of control $u_k$ at time $k$ from the current state $x_k$, and $g_k (x_k, u_k, \eta, \rho, \delta)$ is the stage cost.
A policy $\pi = \{ \mu_1, \dots, \mu_{N}\}$ is a list of functions, where $\mu_k$ maps state $x_k$ into control $u_k = \mu_k(x_k) \in \mathbf{U}_k(x_k)$.
The cost of policy $\pi$ at state $x_1$ is expressed as 
$
    J_\pi(x_1) = E \left[ g_N(x_N) + \sum_{k=1}^{N-1} g_k(x_k, \mu_k(x_k), \eta, \rho, \delta)\right],
$
where $g_N(x_N)$ is the terminal cost.
The Bellman equation for the optimal policy $\pi^*=\{\mu_1^*, \dots, \mu_{N}^*\}$ is
\vspace{-5pt}
\begin{equation}
 \mu_k^*(x_k) \in \argmin_{u_k \in \mathbf{U}_k(x_k)} E[ g_k(x_k, u_k, \eta, \rho, \delta) +
     J_{k+1}^*(x_{k+1})],
     \label{bellman_equation}
\end{equation}
$k=1,\dots,N$. We set the stage cost 
as the number of outstanding requests or $g_k (x_k, u_k, \eta, \rho, \delta)= |\mathbf{\Bar{r}_k}|$.
The set of outstanding requests, $\mathbf{\Bar{r}_k}$, at time $k$ depends on the current control input $u_k$ and the random variable $\eta$. We set $|\mathbf{\Bar{r}_k}| = |\mathbf{\Bar{r}_{k-1}}| + \eta(k) - \psi(x_k, u_k)$, where the function $\psi(x_k, u_k)$ determines the number of requests serviced by executing control $u_k$ at state $x_k$. Since there is no request before $k=1$, we set $|\mathbf{\bar{r}_{0}}| = 0$, $\psi(x_1, u_1) = 0$ and $|\mathbf{\Bar{r}_1}| = |\mathbf{r_1}|$. The optimal cost for our taxicab problem is, 
    $J_{\pi^*}(x_1) = \min_{\pi \in \Pi} E \left[ \sum_{k=1}^{N} g_k (x_k, u_k, \eta, \rho, \delta) \text{ }\big| \text{ } \pi, \eta,\rho,\delta\right].$

The size of the state space at a time $k$, with $m$ available taxicabs is $O({|\mathds{V}|}^m ({|\mathds{V}|\times |\mathds{V}|})^{|\mathbf{r_k}|})$ since each available taxicab can be located at any of the $|\mathds{V}|$ locations and there may be $|\mathds{V}|\times |\mathds{V}|$ possible pickup-dropoff location pairs for each request. The control space grows exponentially with the number of agents. Finding an optimal policy for such a large multiagent taxicab routing problem is intractable, and hence, we look for suboptimal solutions.

\section{Our Approach}\label{sec:our_approach}
Our approach can be characterized as a form of approximation in value space, built on top of a self-learning policy iteration scheme. It has: a) {\it An online play algorithm} that leverages the results of the offline policy iteration. b) {\it An offline training algorithm} that is based on approximate policy iteration, with the policy cost functions approximated using neural networks. c) {\it Agent-by-agent rollout policies} for faster generation of training samples for policy evaluation.

Through online play as an approximate policy improvement step, we obtain a competitive suboptimal policy that improves over the performance of the offline policy (see Sec.~\ref{subsec_preliminary}). 
We choose an offline policy that approximates a variant of the rollout algorithm known as one-agent-at-a-time (one-at-a-time) rollout~\cite{BERTSEKAS2020Multiagent} \cite{Bertsekas2021PI}, as it scales linearly with the number of agents, instead of exponentially like the standard rollout algorithm \cite{bertsekas2019reinforcement}. We implement the offline approximation using GNNs to leverage the topological structure of the street network (see Sec.~\ref{subsec_offline_archi}). 
Since our policy approximation captures the behavior of a rollout policy derived for a specific historical demand model, a deviation in the current demand model from the original historical demand model might affect the quality of the approximation. This change in the approximation performance will in turn affect the performance of the online play. 
To address this, we quantify regions of validity using Wasserstein Ambiguity Sets \cite{Esfahani2015}, and propose a method for switching to a different offline approximation based on the Wasserstein Distance \cite{villani2009optimal} \cite{Ran2020} between the current and the historical demand model used for training the offline approximation (see Sec.~\ref{subsec_adaptivity}).

\subsection{Background: rollout algorithm, offline approximation, and online play}\label{subsec_preliminary}

\subsubsection{One-agent-at-a-time (one-at-a-time) rollout}
We now discuss rollout \cite{bertsekas2020rollout},\cite{Bhattacharya2020MultiagentRollout}, denoted by $\tilde{\pi}$, where the optimal cost $J^*_{k+1}$ in the Bellman equation (Eq.~\ref{bellman_equation}) is replaced with a cost approximation $\Tilde{J}_{k+1}$. This rollout finds a non-myopic solution with consideration of the future using lookahead optimization.
Each agent's control is obtained by performing one-step lookahead minimization over the agent's control components. Rollout's exhaustive expectation estimation performs better than other RL algorithms (including MCTS~\cite{SiV10}) that use longer and sparser lookahead trees with inexact expectation estimation~\cite{Bhattacharya2020RAL}.
The cost approximation $\tilde{J}$ can be estimated by the cost of applying a base policy $\pi$, for $t$ times followed by a terminal cost approximation $\hat{J}$. A base policy can be given by simple heuristics.
Agent $\ell$'s one-at-a-time rollout control at state $x_k$, is
\begin{equation}
    \Tilde{u}_k^{\ell} \in   \argmin_{u^\ell_k \in \mathbf{U}^\ell_k(x_k)} E[ g_k(x_k, {\bar{u}}, \eta, \rho, \delta) 
    +\Tilde{J}_{k+1}(x_{k+1})]
     \label{eq:OAT-rollout}
\end{equation}
where ${\bar{u}}=(\Tilde{u}_k^{1},\ldots, \Tilde{u}_k^{\ell-1}, u^\ell_k, \mu_k^{\ell+1}(x_k),\ldots, \mu_k^{m}(x_k))$. This agent by agent optimization scales linearly with the number of agents.
\cite{BERTSEKAS2020Multiagent}, \cite{Bertsekas2021PI}, and \cite{bertsekas2020rollout} show that the one-at-a-time rollout with one-step lookahead guarantees \textit{cost improvement} \cite{bertsekas2020rollout} of the rollout policy ${\tilde{\pi}}$, improving over base policy ${\pi}$.

\subsubsection{Offline Approximation and Online Play}\label{subsec_offline_approx_online_play}
\begin{figure}[ht]
        \centering
        \vspace{-5pt}
        \includegraphics[width=0.45\textwidth]{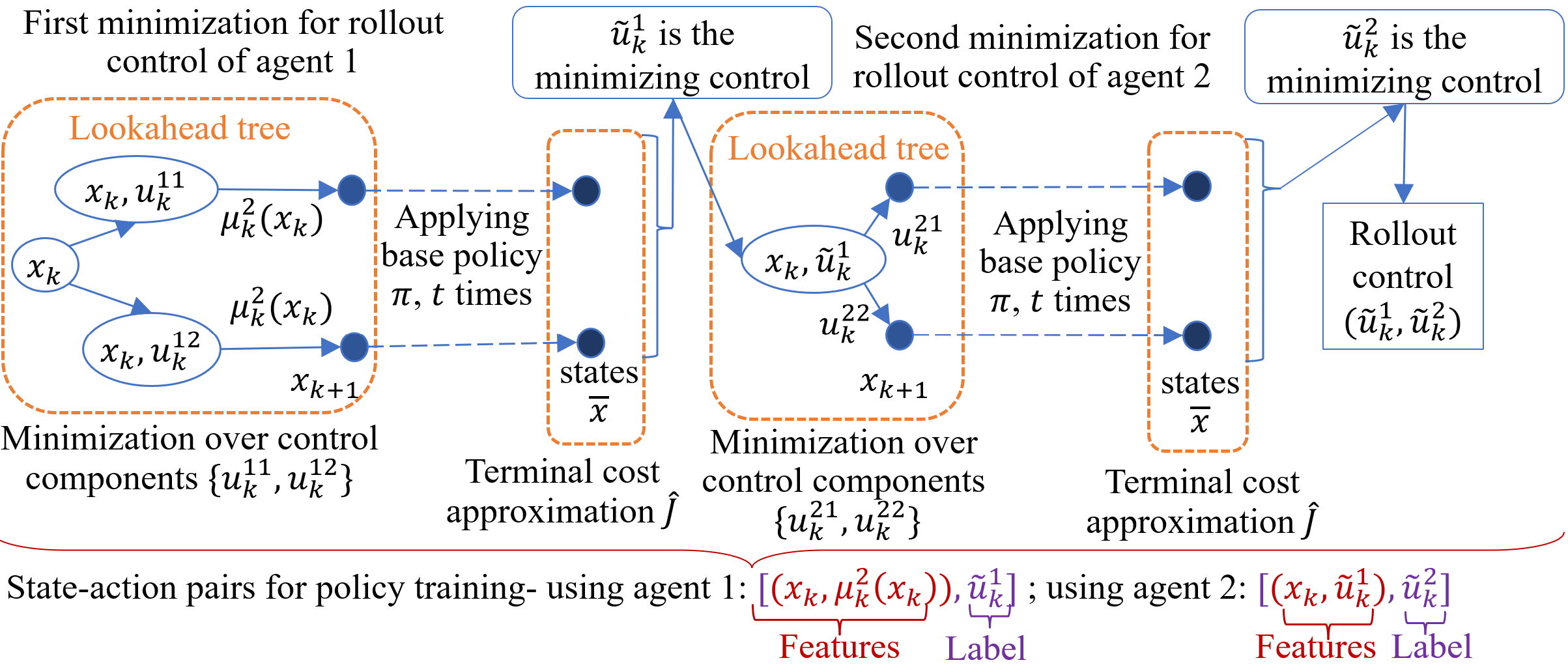}
        \vspace{-10pt}
        \caption{\small{Our pipeline for training data generation for the offline policy approximation of one-at-a-time rollout with two agents.
        }}
        \vspace{-8pt}
        \label{fig:training_data}
    \end{figure}

Now, we discuss the offline trained policy $\hat{\mu}$
used to approximate the one-at-a-time rollout-based RL policy. 
To train the policy approximation, we generate a large set of random states, each with random initial taxicab locations. The state and other agent's actions are features, while the corresponding one-at-a-time rollout controls are labels.
The training feature at state $x_k$ for agent $\ell$ is $F(x_k, \ell) = (x_k,\Tilde{u}_k^{1},\ldots, \Tilde{u}_k^{\ell-1}, \mu_k^{\ell+1}(x_k),\ldots, \mu_k^{m}(x_k))$ and the training label is $\tilde{\mu}^\ell_k, \ell\in \{1,\ldots,m\}$. $\hat{\mu}$ is a function that maps $F(x_k, \ell)$ to the rollout control for agent $\ell$.
Fig.~\ref{fig:training_data} shows the training data generation for $2$ agents.

The online play\cite{Bertsekas2022AlphaZero}, denoted by $\bar{\pi}$, evaluates the one-at-a-time rollout policy with the policy approximation $\hat{\pi}$ as base policy which makes online play act as an approximate policy iteration step over the rollout policy $\tilde{\pi}$.
Online play's control $\bar{u}_k^{\ell}$ at state $x_k$ is given by Eq.~\ref{eq:OAT-rollout}, with ${\bar{u}}=(\bar{u}_k^{1},\ldots, \bar{u}_k^{\ell-1}, u^\ell_k, \hat{\mu}(F(x_k,\ell+1)),\ldots, \hat{\mu}(F(x_k,m)))$.

If the policy approximation, $\hat{\mu}$, correctly approximates the rollout policy, $\tilde{\pi}$, on the current demand model, we expect online play policy $\Bar{\pi}$ to outperform $\Tilde{\pi}$ following the cost improvement property of the approximate policy iteration \cite{bertsekas2019reinforcement}. If $\hat{\mu}$ fails to approximate the rollout policy $\tilde{\pi}$ because of a change in the current demand model, the cost improvement property will not hold and the online play will not provide a significant improvement.
We address these issues in our discussion of adaptivity in Sec.~\ref{subsec_adaptivity}. Next, we present our approach for policy approximation.

\subsection{A hybrid framework for offline approximation and online play for policy generation in the taxi routing problem} \label{subsec_offline_archi}
\begin{figure}[ht]
        \centering
        \vspace{-10pt}
        \includegraphics[width=0.47\textwidth]{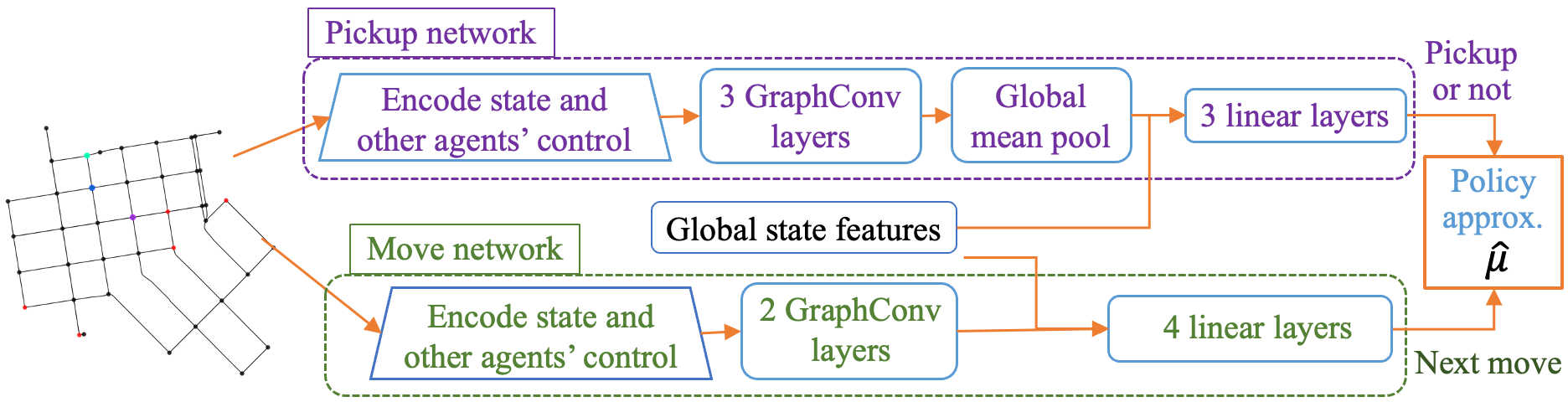}
        \vspace{-5pt}
        \caption{\small{Graph convolutional neural network architectures for the offline policy approximation $\bar{\pi}$. 
        }}
        \vspace{-10pt}
        \label{fig:offline_approximation}
    \end{figure}

Now, we present our approach of approximating the one-at-a-time rollout policy using GNNs (implemented as suggested in \cite{Kipf2016}).
Since we encode the environment as a graph, using GNNs allows us to leverage the connectivity between intersections to boost the performance of the approximation. In our method, we use GNNs \cite{Scarselli2009} as the offline approximation to approximate the one-at-a-time rollout with base policy $\pi$. 
Since encoding the behavior of one-at-a-time rollout is a complex task composed of two main actions, pickup and movement, we separate the offline approximation into two networks. The first GNN determines if an available agent should pickup a request in its current location, and the second GNN determines the next intersection towards which the agent should move. If the first GNN determines that an agent should pick a request, the output of the second GNN is ignored. The pickup GNN is composed of $3$ graph convolutional layers, followed by $3$ linear layers, while the move GNN is composed of $2$ graph convolutional layers, followed by $4$ linear layers (see Fig.~\ref{fig:offline_approximation}). We select these architecture parameters after performing a hyperparameter search. We train a pair of GNNs for all agents for each representative demand model. 
The training data for the GNN is generated following the process shown in Fig.~\ref{fig:training_data}. We encode the state $x_k$ as a set of node features and global features. 
Global features ($\in\mathbb{R}^m$) describe global properties of the state, containing only the list of time remaining $\boldsymbol{\tau_k}$ in current trips for all agents.
Node features ($\in\mathbb{R}^{m+2}$), at each intersection, encode the presence of 
each of $m$ agents, 
indicate if the intersection is chosen as the next move for any of the other agent's potential current actions, 
and include the number of pickup requests available at the intersection.

With this implementation of the offline approximation, we apply online play (given in Sec.~\ref{subsec_offline_approx_online_play}). Next, we explain how we deal with fluctuations in the current demand model.

\subsection{Adaptivity to changing demand distributions}\label{subsec_adaptivity}
We propose a method for providing adaptivity to our GNN-based policy approximation by quantifying regions of validity. These regions are defined using Wasserstein Ambiguity Sets which rely on the Wasserstein Distance, a distance function between two probability distributions on a given probability space. 
We choose the Wasserstein Distance metric because (i) this metric considers the closeness between support points while other metrics only consider their probabilities, and (ii) Wasserstein Ambiguity Sets are rich enough to consider discrete distributions outside of the original support while other metrics, including Kullback-Leibler divergence, do not allow for this.

We now formally define the Wasserstein distance. For this, we choose the estimated distribution for the number of requests, $\Tilde{p}_\eta$, to be the reference distribution, since the variation in $\Tilde{p}_\eta$ over the peak and non-peak hours is most significant.
However, the full quantification of the distance between demand models may include the pickup ($\Tilde{p}_\rho$) and dropoff ($\Tilde{p}_\delta$) distributions, which change less dramatically over the hours of a day. In this setting, we define an atom as a measurable set which has a positive measure and contains no set of smaller positive measure~\cite{Azram2011}.
For a finite support  $\Omega$, which includes $X$ data points, let $\xi_j^c$ denote the $j$-th atom of the current demand distribution $\Tilde{p}_{\eta,c}$, and $\xi_i$ be the $i$-th atom of the representative historical distribution $\Tilde{p}_\eta$. 
$p_j^c$ denotes the probability of $\xi_j^c$, 
$p_i$ denotes the probability of $\xi_i$,
and $f_{ij}$ denotes the bivariate probability mass function for $\xi_j^c$ and $\xi_i$,  $\forall i,j \in \{1,\ldots,X\}$. 
With these definitions, the Wasserstein Distance~\cite{villani2009optimal} $d_W(\Tilde{p}_{\eta,c}, \Tilde{p}_\eta)$ of order 1, also known as the Kantorovich distance, is defined as
$d_W(\Tilde{p}_{\eta,c}, \Tilde{p}_\eta) = {\textstyle\inf}_{f \geq 0} {\textstyle\sum}_{i,j \in \{1,\ldots,X\}} f_{ij} ||\xi_j^c - \xi_i ||$,
subject to ${\textstyle\sum}_{j \in \{1,\ldots,X\}} f_{ij} = p_i, \forall i \in \{1,\ldots,X\}$, and ${\textstyle\sum}_{i \in \{1,\ldots,X\}} f_{ij} = p_j^c, \forall j \in \{1,\ldots,X\}$. 

We now formally define the Wasserstein Ambiguity Set and $q$-valid radius.
We let $\mathcal{P}(\Omega)$ represent the space of all underlying probability distributions $p_{\eta,c}$ supported on $\Omega$. The Wasserstein Ambiguity Set, $\mathcal{D}_W$, is a ball of radius $\theta$ centered around the reference distribution $\Tilde{p}_\eta$ defined as $ \mathcal{D}_W := \{ p_{\eta,c} \in  \mathcal{P}(\Omega) | d_W(p_{\eta,c}, \Tilde{p}_\eta) < \theta\}$ \cite{Esfahani2015}. We choose the radius $\theta$ for the ambiguity sets to be the lower bound for the $q$-valid radius as in Theorem 2 of \cite{Ran2020} 
$$\theta \geq ( B + 0.75) ( - \log(1-q)/X + 2 \sqrt{- \log(1-q)/X} ),\vspace{-5pt}$$
where the $q$-valid radius, following ~\cite{Ran2020}, corresponds to the radius that ensures $\Tilde{p}_{\eta, c} \in \mathcal{D}_W$ with probability at least $q$, defining the bounds for the region of validity; $B$ is the diameter of the 
compact support $\Omega$, which in the taxicab routing setting is the maximum number of requests/minute.
Following this formulation, if the current demand distribution is outside the $q$-valid radius for a given representative historical distribution, then there is a different representative distribution that contains the current demand distribution inside its ambiguity set with a higher probability. Therefore, whenever the current demand model gets outside of the $q$-valid radius, we choose the policy approximation, trained on the representative demand closest to the current demand in Wasserstein distance.
\begin{figure}[ht]
        \centering
        \vspace{5pt}
        \includegraphics[width=0.38\textwidth]{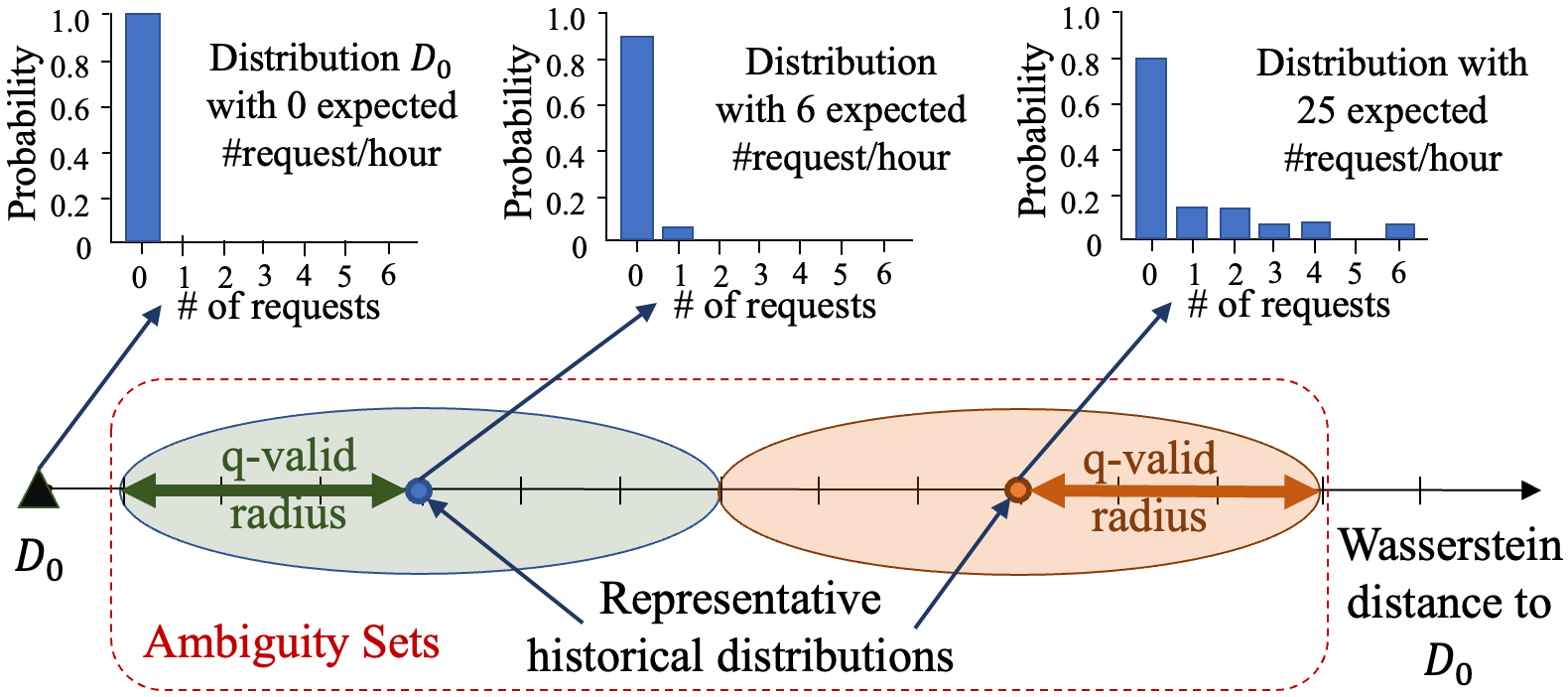}
        \caption{\small{Ambiguity sets induced by $q$-valid radii of the representative distributions from the historical demand model.
        }}
        \vspace{-10pt}
        \label{fig:ambiguity_sets}
    \end{figure}

\section{Numerical Experiments}
\label{sec:result}
Now, we outline our implementation details and simulation results on a real taxicab dataset~\cite{epfl-mobility-20090224}. We show that our method outperforms rollout and several OR based benchmarks, being robust towards changing demands. 

\subsection{Experimental Setup}
\label{subsec:experimental_setup}
We consider a $400\times400 m^2$ section of San Francisco's financial district~\cite{epfl-mobility-20090224} (see Fig.~\ref{fig:graph/map}) with $42$ nodes and $125$ edges, and $m=3$ taxicabs (with the state and control space sizes of $10^{78}$, and $216$, respectively).
We consider $1$-minute edge travel time and the time horizon of $N=60$ minutes.

\subsection{Estimating the demand model}\label{sec:demand_model}
\begin{figure}[ht]
        \vspace{-8pt}
        \centering
        \includegraphics[width=0.35\textwidth]{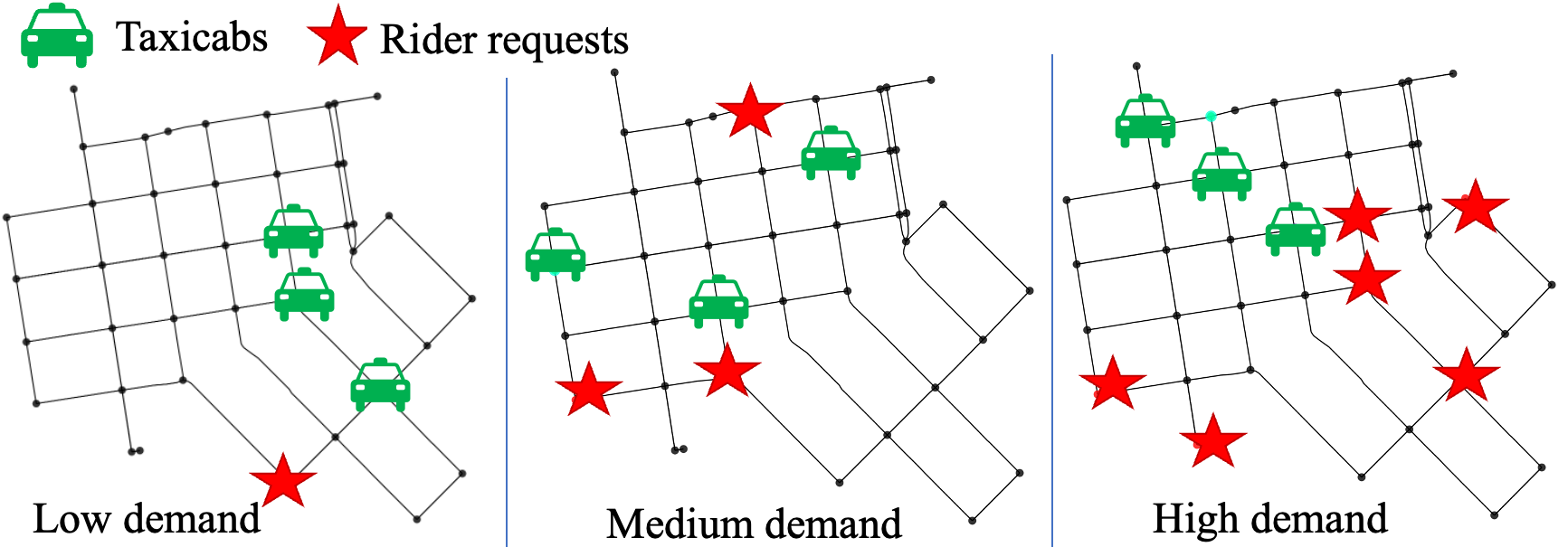}
        \vspace{-8pt}
        \caption{\small{Visualizing various demand models.
        }}
        \vspace{-8pt}
        \label{fig:visual_load}
    \end{figure}

The historical demand model is composed of three estimated categorical distributions $\Tilde{p}_\eta$ for number of requests, $\Tilde{p}_\rho$ for pickup locations, and $\Tilde{p}_\delta$ for dropoff locations. We partition the historical data in 1-hour intervals, where each time step $k$ spans 1 minute. We empirically estimate $\Tilde{p}_\eta$ by looking at the number of requests that arrive at each minute within each 1-hour time span. $\Tilde{p}_\rho$ and $\Tilde{p}_\delta$ are derived from the historical requests \cite{epfl-mobility-20090224} that originated and ended inside the map section. The probability of a request emerging at pickup node $y$ is $\Tilde{p}_{\rho}(y) = (s_y +1/|\mathds{V}|)/(1+ {\textstyle\sum}_{j \in \mathds{V}} s_j),$
where $s_y$ is the number of requests with pickup location $y$.
We assign a small nonzero probability of request origination to all intersections to represent the idea that requests may originate at any intersection.
The dropoff location probability $\Tilde{p}_\delta$ is estimated similarly from 
the historical dropoffs in \cite{epfl-mobility-20090224}. For our experiments, we consider three different demand models: low, medium, and high, (see Fig.~\ref{fig:visual_load}) with the same $\tilde{p}_{\rho}$ and $\tilde{p}_{\delta}$, but with different $\tilde{p}_{\eta}$. 
The low, medium and high demand models have $E[\eta] \cdot N$ of $3,9$ and $25$, respectively.

The current demand model ($\Tilde{p}_{\eta,c}, \Tilde{p}_{\rho,c}, \Tilde{p}_{\delta,c}$) is derived similarly by looking at the latest hour of execution of the system, instead of looking at historical data. The historical demand models are used to train the policy approximation, while the current demand model is used as the demand model for evaluation of all policies.

\subsection{Implementation details of one-at-a-time rollout}\label{subsec:oat}
We consider a greedy base policy $\pi$ where taxis are routed to their nearest request without coordination. The one-at-a-time rollout employs a 1-step lookahead optimization tree at stage $k$, where each leaf node estimates the cost using $5000$ Monte-Carlo simulated trajectories. Each trajectory applies the greedy base policy $t=10$ times before truncation.
We set the terminal cost approximation $\hat{J}=|\mathbf{\Bar{r}}_{k+t+1}|$; the remaining number of outstanding requests at the time of truncation.

\subsection{GNN offline policy approximation architecture}\label{subsec:GNN}
We use the following setup to train the GNN architectures.
We use Adam optimizer \cite{Kingma2014AdamAM} with a learning rate of $0.005$ and $0.002$ for pickup and move networks, respectively, and a regularizing factor of $ 10^{-5}$ for generalization. We use $1.2\times 10^6$ random state-(rollout) control pairs and $100$ epochs. It took $16$ hours to train each move network and $6$ hours to train each pickup network on a single NVIDIA RTX A6000.

\subsection{Parameters for Ambiguity Sets}
To demonstrate adaptivity of our method, we introduce variability in the current demand model for $p_\eta$. We set the diameter $B$ of $\Tilde{p}_\eta$ as 6 (since there are between 0 and 6 requests per minute for the map section in Fig.~\ref{fig:graph/map} for the San Francisco taxicab dataset \cite{epfl-mobility-20090224}). Since we use $5000$ samples from the historical distribution to approximate the expectation during training of the offline approximation, we set $X=5000$. Using these values, we set $q=0.54$ and obtain that the minimum $q$-valid radius is $0.114$. We choose $q$ to be a little higher than $50\%$ to guarantee that the sets are as small as possible, while still having a higher probability of containing the current demand distribution inside the set. We choose representative historical distributions that are centered around expected demand values with high relative frequencies in the historical data (See Fig.~\ref{fig:ambiguity_sets} for an example).

\subsection{Benchmarks}\label{subsection_benchmarks}
In this section, we discuss benchmarks from relevant OR methods for our comparison study.

\textbf{Instantaneous assignment:} This algorithm, inspired by BLE~\cite{werger2000broadcast} for iterative task assignment, performs a deterministic matching of outstanding requests and available taxis. This method does not consider future requests.

\textbf{Two-step stochastic optimization (TSS):} This multiagent task assignment algorithm \cite{LOWALEKAR201871} performs a maximization of the combined reward (negative wait time) of assigning taxicabs to requests using the set of currently outstanding requests and possible requests at the next stage predicted using the current demand model. We use $1000$ sets of request samples to estimate the requests at the next stage.

\textbf{Oracle:} This method~\cite{LOWALEKAR201871} solves an optimal multiagent task assignment using the \textit{full knowledge of the current and all future requests} in the planning horizon by maximizing the combined reward (negative wait time) of all assignments. This method is only used as 
a lower bound on the cost assuming we had full a-priori information about the pickup/dropoff locations and request arrival times for all requests.

\subsection{Numerical performance for taxicab pickup problem}\label{sec:simu_result}
Now, we present a comparative study of online play with the benchmark methods, and different components of online play, including greedy policy and one-at-a-time rollout from Sec.~\ref{subsec:oat}, and GNN policy approximation from Sec.~\ref{subsec:GNN}. All results in this section use average wait time over $50$ random starting states, and they are expressed using min-max normalization. This normalization transforms the average total wait time of a policy $\pi$ over all evaluation states at a given demand from $\bar{J}_\pi$ minutes to $\bar{J}^{norm}_\pi\in[0,1]$, where 
$\bar{J}^{norm}_\pi=(\bar{J}_\pi-min_{\pi'\in \Pi_e} \bar{J}_{\pi'})/(max_{\pi'\in \Pi_e}\bar{J}_{\pi'}-min_{\pi'\in \Pi_e}\bar{J}_{\pi'}).$
Here, $\Pi_e \subseteq$ \{Greedy policy, rollout, GNN, online play, instantaneous assignment, TSS, oracle\}.

First, we are interested in the performance of online play when the current demand agrees with the historical demand used for training the offline approximation. Table~\ref{tab:claim1} shows that online play outperforms 
all the other methods. 

\begin{table}[ht]
           \centering
           \vspace{-8pt}
           \caption{Normalized wait time  with GNN approximation trained on the same demand distribution used for evaluation}
           \vspace{-8pt}
           \resizebox{0.8\linewidth}{!}{
               \begin{tabular} { |c|c|c|c|}
                       \hline
                        \multirow{2}{*}{Policies} & \multicolumn{3}{c}{Normalized wait time for demand models: }
                   \\ 
                    & \makecell{Low (L)} & \makecell{Medium (M)} & \makecell{High  (H)}
                   \\ \hline 
Greedy policy & 0.94 & 0.99 & 1.0\\ \hline 
One-at-a-time rollout & 0.62 & 0.65 & 0.58\\ \hline 
GNN & 0.58 & 0.68 & 0.65\\ \hline 
Online play w. GNN & \textbf{0.57} & \textbf{0.62} & \textbf{0.5}\\ \hline 
Inst. assign. & 1.0 & 0.92 & 0.88\\ \hline 
TSS & 0.96 & 1.0 & 0.89\\ \hline 
Oracle & 0.0 & 0.0 & 0.0\\ \hline 
\makecell{Min/Max for\\normalization (in minutes)} & \makecell{2.5/\\13.0} & \makecell{8.9/\\42.4} & \makecell{219.1/\\276.5}\\ \hline 

                   \end{tabular}}
                   \label{tab:claim1}
                   \vspace{-10pt}
\end{table}

Second, we empirically show the robustness and limitations of the online play with an offline trained approximation by evaluating on out-of-distribution (OOD) demand. OOD demand corresponds to the case when the current demand model deviates from the historical demand model used for training. In Table~\ref{tab:claim2_1}, we show that online play outperforms all the other methods for distributions that are inside the $q$-valid radius ($0.114$) of its Wasserstein Ambiguity Set. We also show the eventual performance degradation of online play for distributions outside the $q$-valid radius.

\newcolumntype{g}{>{\columncolor{green}}c}
\begin{table}[ht]
            \vspace{-5pt}
            \centering
            \caption{Normalized wait time 
            for different demand distributions
            }
            \vspace{-8pt}
            \resizebox{0.88\linewidth}{!}{
                \begin{tabular} { |c|g|g|g|c|c|c|}
    \hline \multirow{3}{*}{Policies} & \multicolumn{6}{c}{Wasserstein Distance between training and evaluation model}\\
    & \multicolumn{3}{c}{Within q-valid radius} & \multicolumn{3}{c}{Outside q-valid radius} \\
      & 0.0 & 0.017 & 0.067 & 0.117 & 0.15 & 0.35\\ \hline 
Greedy policy & 1.0 & 1.0 & 1.0 & 1.0 & 1.0 & 0.77\\ \hline 
\makecell{One-at-a-time\\rollout} & 0.66 & 0.58 & 0.64 & \textbf{0.62} & \textbf{0.6} & \textbf{0.46}\\ \hline 
\makecell{GNN trained\\on low demand}& 0.62 & 0.62 & 0.73 & 0.74 & 0.98 & 1.0\\ \hline 
\makecell{Online play w.\\GNN trained\\on low demand}& \textbf{0.61} & \textbf{0.57} & \textbf{0.63} & 0.73 & 0.68 & 0.52\\ \hline 
Oracle & 0.0 & 0.0 & 0.0 & 0.0 & 0.0 & 0.0\\ \hline 
\makecell{Min/Max for\\normalization \\(in minutes) } & \makecell{2.5 / \\ 12.4} & \makecell{2.2 / \\ 16.7} & \makecell{3.7 / \\ 29.4} & \makecell{15.3 / \\ 55.3} & \makecell{20.4 / \\ 62.6} & \makecell{201.3 / \\ 289.9}\\ \hline 
                    \end{tabular}}
                    \label{tab:claim2_1}
                    \vspace{-10pt}
            \end{table}

\begin{figure}[ht]
        \centering
        \vspace{5pt}
        \includegraphics[width=0.38\textwidth]{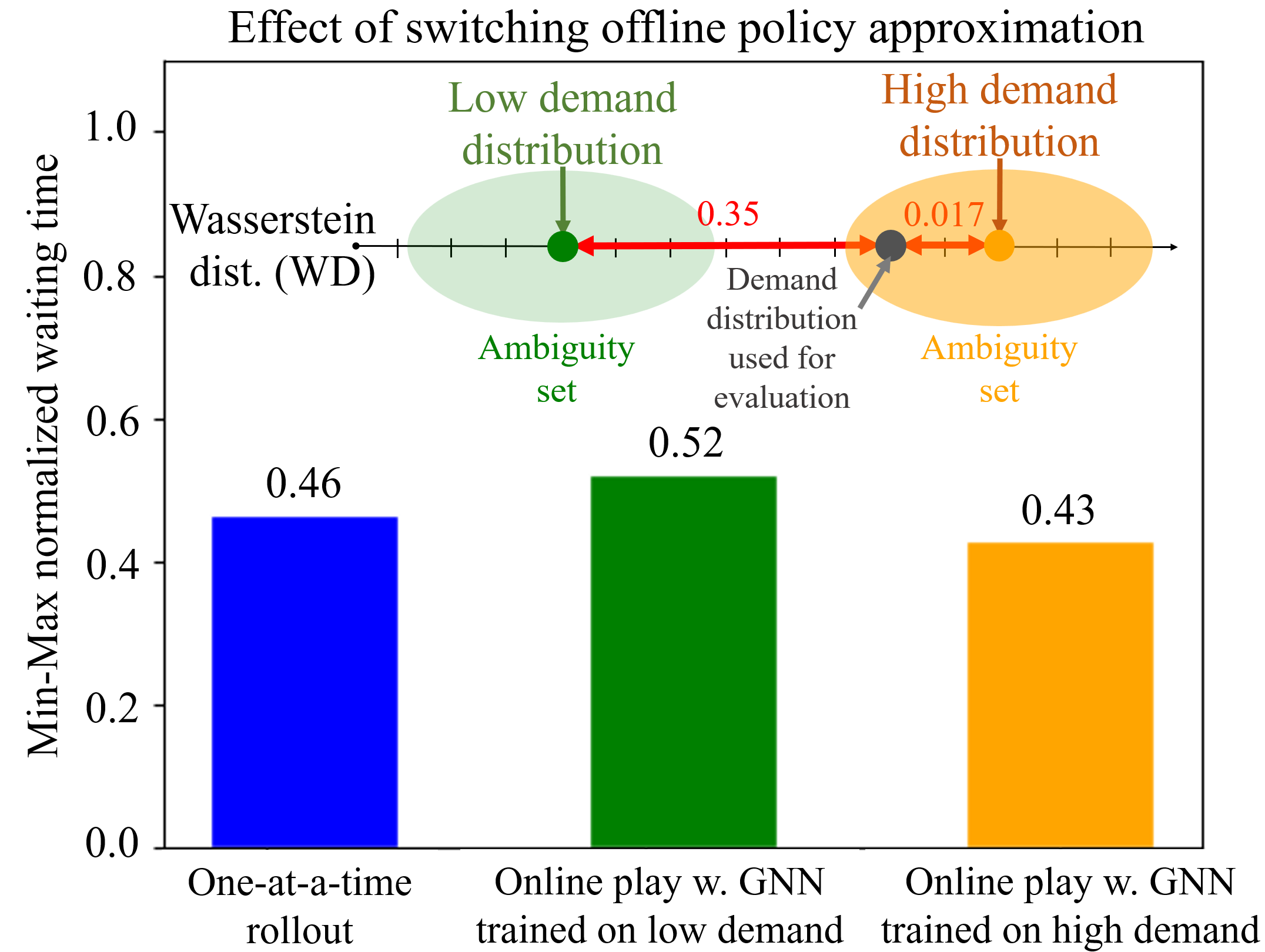}
        \vspace{-8pt}
        \caption{\small{Performance of online play evaluated 
        on a distribution that lies inside of high demand ambiguity set and outside of low demand ambiguity set.
        The minimum and maximum values used for normalization are $201.6$ and $289.9$, respectively.}}
    \vspace{-15pt}
        \label{fig:results_relative_performance}
    \end{figure}

Once the current demand model is outside the $q$-valid radius of the historical demand model, our approach recovers the performance gain over rollout by exchanging the original policy approximation for one trained on a distribution that contains the current demand model inside its ambiguity set.
Fig.~\ref{fig:results_relative_performance} shows that switching to a better policy approximation gives a lower cost for online play ($9\%$ relative improvement).

\subsection{Scalability}\label{sec:scalability}
In this section, we present the numerical results in a map of $1500\times1500 m^2$ ($825$ nodes and $1884$ edges) with $m=15$ agents (with a state-space size of $10^{459}$, and a control-space size of $10^{12}$). Increasing the map size linearly increases the size of the input feature for the GNN approximation. Efficiently tackling this curse of dimensionality prove to be a hard task, and hence the focus of future work. 

The online optimization aspect of our approach can be used for bigger setups since rollout does online replanning automatically and does not rely on the GNN approximation. To enable our approach to scale up, we consider the same one-agent-at-a-time one-step lookahead optimization (proposed in \cite{BERTSEKAS2020Multiagent} \cite{Bertsekas2021PI}), with the lookahead step being fully stochastic. We estimate the cost approximation of the Q-factors at each leaf node of the lookahead tree by applying truncated rollout with a Certainty Equivalence \cite{BertsekasCE} approximation.  We use instantaneous assignment using an auction algorithm as suggested in \cite{Bertsekas1979Auction} \cite{bertsekas1998network} \cite{Bertsekas2020Auction} as the base policy, allowing our online optimization scheme to leverage the matching solution as the starting point for the optimization. In our implementation of the Certainty Equivalence approximation, we fix the disturbances $\eta$, $\rho$, and $\delta$ across all the rollout steps, only preserving the stochasticity of the order in which requests arrive and the pairing between pickup and dropoff locations for each request. Reducing uncertainty in disturbances enables us to use fewer ($2000$) Monte-Carlo simulations per leaf node.

Table~\ref{tab:scalability} shows normalized results averaged over $50$ trajectories. For our experiments, we evaluate our method on three different demand models: low, medium, and high, with the same $\tilde{p}_{\rho, c}$ and $\tilde{p}_{\delta, c}$, but with different $\tilde{p}_{\eta,c}$. The low, medium and high demand models have $E[\eta] \cdot N$ of $15,45$ and $75$, respectively. Table~\ref{tab:scalability} shows that our approach outperforms the greedy policy and the instantaneous assignment baseline. We do not include results for TSS in this larger map, due to its prohibitively long runtime.

\begin{table}[ht]
           \centering
           \vspace{-8pt}
           \caption{Min/Max normalized wait time for bigger setup}
           \vspace{-8pt}
           \resizebox{0.88\linewidth}{!}{
               \begin{tabular} { |c|c|c|c|c|c|}
                       \hline
                        \multirow{2}{*}{Demand} & \multicolumn{4}{c}{Policies} & \multirow{2}{*}{Min / Max}
                   \\ 
                    & \makecell{Greedy} & \makecell{Inst. assign.} & \makecell{Our method} &\makecell{Oracle}
                   \\ \hline 
Low & 1.0 & 0.86 & \textbf{0.77} & 0.0 & 9.9 / 137.2\\ \hline
Medium & 1.0 & 0.86 & \textbf{0.83} & 0.0 & 207.1 / 502.3\\ \hline 
High & 1.0 & 0.99 & \textbf{0.89} & 0.0 & 684.4 / 995.4\\ \hline 
                   \end{tabular}}
                   \label{tab:scalability}
                   \vspace{-10pt}
\end{table}

\section{Conclusion}
In this paper, we apply online play in combination with offline policy approximations and verify that our approach allows the system to adapt to changes in the underlying demand conditions. Our future work includes, but is not limited to, considerations of travel time between intersections, time allocations for servicing vehicles, and predicting future demand to switch policy approximations proactively.







\bibliographystyle{IEEEtran}
\bibliography{IEEEabrv, main}

\end{document}